\documentclass[]{pasj01}

\begin{document} 
\Received{}
\Accepted{}

\title{Revised Wavelength and Spectral Response Calibrations for AKARI Near-Infrared Grism Spectroscopy: Post-Cryogenic Phase}

\author{Shunsuke \textsc{baba}\altaffilmark{1}}
\altaffiltext{1}{Institute of Space and Astronautical Science, Japan Aerospace Exploration Agency, 3-1-1 Yoshinodai, Chuo-ku, Sagamihara, Kanagawa 252-5210, Japan}
\email{s-baba@ir.isas.jaxa.jp}

\author{Takao \textsc{nakagawa}\altaffilmark{1}}

\author{Fumihiko \textsc{usui}\altaffilmark{2}}
\altaffiltext{2}{Center of Planetary Science, Graduate School of Science, Kobe University, 7-1-48 Minatojima-Minatomachi, Chuo-ku, Kobe, Hyogo 650-0047, Japan}

\author{Mitsuyoshi \textsc{yamagishi}\altaffilmark{1}}

\author{Takashi \textsc{Onaka}\altaffilmark{3}}
\altaffiltext{3}{Department of Astronomy, Graduate School of Science, The University of Tokyo, 7-3-1 Hongo, Bunkyo-ku, Tokyo 113-0033, Japan}



\maketitle

\begin{abstract}
We present a new calibration for the second-order light contamination in the near-infrared grism spectroscopy with the Infrared Camera aboard AKARI, specifically for the post-cryogenic phase of the satellite (Phase 3).
Following our previous work on the cryogenic phase (Phases 1 and 2), the wavelength and spectral response calibrations were revised.
Unlike Phases 1 and 2, during Phase 3, the temperature of the instrument was not stable and gradually increased from 40 to 47~K.
To assess the effect of the temperature increase, we divided Phase 3 into three sub-phases and performed the calibrations separately.
As in Phases 1 and 2, we confirmed that there was contamination due to the wavelength dependence of the refractive index of the grism material in every sub-phase.
The wavelength calibration curves for the three sub-phases coincided with each other and did not show any significant temperature dependence.
The response decreased with temperature by $\sim10\%$ from the beginning to the end of Phase 3.
We approximated the temperature dependence of the response as a linear relation and derived a correction factor as a function of temperature.
The relative fraction of the second-order light contamination to the first-order light was found to be by 25\% smaller than in Phases 1 and 2.
\end{abstract}

\section{Introduction}

The near-infrared (IR) grism spectroscopy performed in the NIR channel of the Infrared Camera (IRC) aboard the AKARI satellite covers a wavelength range of 2.5--5.0~\micron\ with a spectral resolution of $\sim$120 \citep{MurakamiEtAl_2007,OhyamaEtAl_2007e,OnakaEtAl_2007}.
This wavelength range cannot be observed with ground-based telescopes due to atmospheric absorption and emission, and was not accessible with the Spitzer Space Telescope because that satellite could not observe wavelengths shorter than 5.2~\micron\ \citep{HouckEtAl_2004}.
Although the Infrared Space Observatory (\cite{KesslerEtAl_1996}) was also capable of near-IR spectroscopy of a similar spectral resolution with ISOPHOT-S, its sensitivity was roughly $10^3$ times lower than that of the AKARI grism mode \citep{KlaasEtAl_1997}.
AKARI NIR spectroscopy is, hence, a unique capability, at least before the launch of the James Webb Space Telescope.

The NIR channel employs a 512$\times$412 InSb array with a pixel scale of 1\farcs46 \citep{OnakaEtAl_2007}.
Because the detector itself has sensitivity for wavelengths shorter than 2.5~\micron\ and the grism is of transmission type, an order sorting filter cutting off those wavelengths is coated on the front surface of the grism to avoid spectral overlapping between the diffracted first-order light to be observed and higher orders.
Despite the order-sorting filter, however, the first-order spectrum is contaminated by the second-order light at the longer end of the wavelength coverage.\footnote{Section 6.9.4 of AKARI IRC Data User Manual for Post-Helium (Phase 3) Mission (http://www.ir.isas.jaxa.jp/AKARI/Observation/support/IRC/IDUM/
IRC\_IDUM\_P3\_1.1.pdf).}
This means that the filter has a non-negligible leak or that there is a nonlinearity in the position-wavelength relation.
The anomaly prevents us from accurately calibrating the flux in that part of the spectrum and becomes a serious issue in analyses, such as for the redshifted CO ro-vibrational absorption band \citep{BabaEtAl_2018}.

\authorcite{BabaEtAl_2016} (\yearcite{BabaEtAl_2016}; hereafter Paper I) investigated this problem for the cryogenic phase of the satellite (Phases 1 and 2), during which focal plane instruments were kept at cryogenic temperatures ($\sim$6~K) by liquid helium (LHe) and mechanical cryo-coolers to provide high sensitivity.
The correction carried out in Paper I was based on a wavelength calibration that took into consideration the wavelength dependence of the refractive index of the grism material (germanium; Ge) and a spectral response calibration that simultaneously solved the responses to the first- and second-order components using two types of standard objects of contrastive blue and red colors.

The post-cryogenic phase after the exhaustion of LHe is called Phase 3.
In this phase, the IRC was cooled only by a cryo-cooler and had somewhat degraded sensitivity due to the increase of the operating temperature.
Despite the lower sensitivity, in this period, more observations were conducted in the grism mode than in Phases 1 and 2 because all the observation time in Phase 3 was dedicated to NIR pointed observations, while that in the former phases was also allocated for all-sky surveys and mid- and far-IR observations \citep{MatsuharaEtAl_2005}.
The correction procedure for the contamination in Paper I is basically applicable to Phase 3 as well.
We here, however, have to take into consideration the change of the operating temperature among the phases.
The temperature jumped from $\sim$6 to $>$40~K at the beginning of Phase 3.
Besides, the temperature was not stable during Phase 3 and gradually increased with time.
The absolute sensitivity of the NIR grism mode in Phase 3 decreased roughly by 30\% from that in Phases 1 and 2 \citep{OnakaEtAl_2010}.
We now examine the higher order effects considered in Paper I (the nonlinearity of the wavelength calibration and the second-order light contamination) for possible temperature dependences.

In this paper, we explain the revision of the wavelength and spectral response calibrations of the post-cryogenic AKARI NIR grism spectroscopy.
In section \ref{sec.cal3.sub}, we review the temperature variation during Phase 3.
The new wavelength and spectral response calibrations are given in sections \ref{sec.cal3.wave} and \ref{sec.cal3.res}, respectively, followed by comparisons with those for Phases 1 and 2 in section \ref{sec.cal3.comp}.
Finally, we summarize the result of this paper in section \ref{sec.cal3.sum}.
We refer the reader to Paper I for a general description of the AKARI IRC NIR channel and the flux calibration method using the spectral responses to the first- and second-order components.

\section{Definition of Sub-Phases in Phase 3}
\label{sec.cal3.sub}

\begin{figure}
  \begin{center}
    \includegraphics[width=8cm]{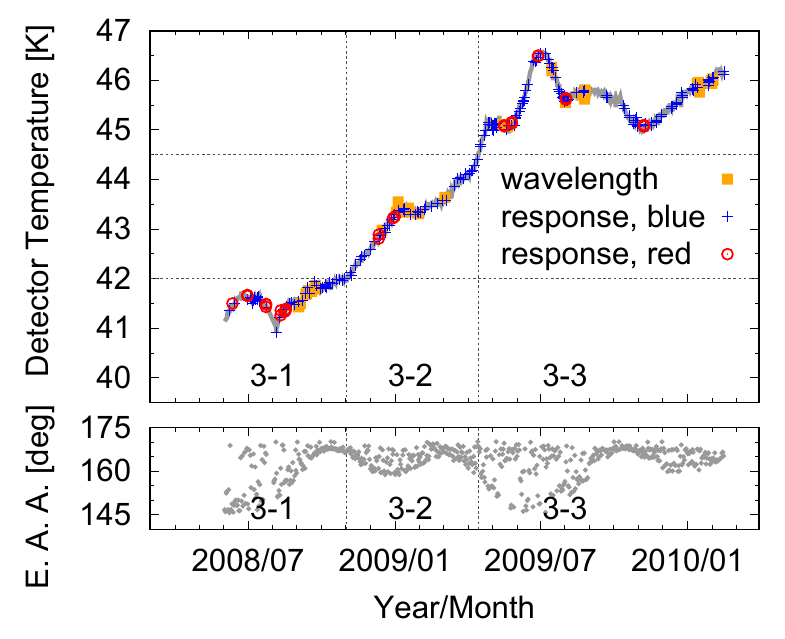}
  \end{center}
  \caption{%
    Top: Detector temperature $T$ in the NIR channel during Phase 3 is shown as a thick gray line.
         The orange filled squares indicate observations used in the wavelength calibration, and the blue crosses and the red open circles represent observations of the blue and red standard objects used in the spectral response calibration.
         The vertical dotted lines are the borders between sub-phases 3-1, 3-2, and 3-3.
         These sub-phases correspond to temperature ranges $T<42~\mathrm{K}$, $42<T<44.5~\mathrm{K}$, and $T>44.5~\mathrm{K}$, as indicated with the horizontal dotted lines.
    Bottom: Earth avoidance angle of the satellite recorded in pointed spectroscopic observations.
            Although the angle changes throughout each observation depending on the orbital location, the mean value during the observation is representatively shown.
            Not all the pointed observations are plotted.
            The number of points is thinned out to about one per day.}
  \label{fig.cal3.temp_time}
\end{figure}

The header of FITS images obtained in AKARI pointed observations contains the detector temperature at the acquisition time for the data.
We collected the temperature from spectroscopic observations carried out in Phase 3.
The obtained temperature variation curve is shown in the top panel of figure \ref{fig.cal3.temp_time}.
The temperature jumped from the value in Phases 1 and 2 ($\sim$6~K) to 41~K because of the depletion of LHe.
The temperature then gradually increased to 46.5~K during the first year, with shoulders at June and December.
In the second year of Phase 3, the detector temperature at first decreased and had a local minimum at the beginning of November 2009.
After that, it rose again until the end of Phase 3.

The seasonal variation of the temperature can be attributed to the change in the Earth avoidance angle, which is here defined as the angle between the telescope axis and the Earth center.
The angle recorded in pointed spectroscopic observations is shown in the bottom panel of figure \ref{fig.cal3.temp_time}.
The AKARI satellite was in a circular Sun-synchronous polar orbit at the altitude of $\sim$700~km and the inclination of 98\fdg2, keeping its telescope direction approximately perpendicular to the Sun and roughly opposite from the Earth \citep{MurakamiEtAl_2007}.
At the solstices, therefore, the Earth avoidance angle became small when the satellite was in a high latitude, which led stronger intensity of earthshine and the increase of the temperature.
At the equinoxes, on the other hand, the angle was almost constant at the highest values, and thus the earthshine was weakest.

The header of the FITS images also records the temperature of the filter wheel, on which the grism was mounted together with a prism for lower-resolution spectroscopy and three filters for imaging.
We found that the temperature of the filter wheel tightly correlated with the detector temperature and was lower than the latter by only 1~K.
Since the temperature dependence of the refractive index of Ge is very small (section \ref{sec.cal3.wave}), we hereafter use the detector temperature as the representative temperature of the NIR channel.

To evaluate the effect of the temperature variation, we divided Phase 3 into three periods as tabulated in table \ref{tab.cal3.def}.
These sub-phases correspond to temperature ranges $T<42~\mathrm{K}$, $42<T<44.5~\mathrm{K}$, and $T>44.5~\mathrm{K}$.
These cuts were determined so that there were a sufficient number of observations for calibration in each sub-phase.
We performed wavelength and spectral response calibrations separately for each sub-phase to assess their temperature dependence, as explained in the following sections.

\begin{table}
  \tbl{Definition of sub-phases and their temperature ranges.}{%
	  \begin{tabular}{ccc}
	    \hline
	    Sub-phase  & Date                       & Detector        \\
	               &                            & temperature (K) \\
	    \hline
	    3-1        & 2008 Jun  1 to 2008 Oct 31 & 40.0--42.0 \\
	    3-2        & 2008 Nov  1 to 2009 Apr 14 & 42.0--44.5 \\
	    3-3        & 2009 Apr 15 to 2010 Feb 15 & 44.5--47.0 \\
	    \hline
	  \end{tabular}
  }
  \label{tab.cal3.def}
\end{table}

\section{Wavelength Calibration}
\label{sec.cal3.wave}

In the wavelength calibration of the official data-reduction pipeline ``IRC Spectroscopy Toolkit for Phase 3 data (version 20150331),''\footnote{Distributed at the AKARI web page (http://www.ir.isas.jaxa.jp/AKARI/ Observation/support/IRC/).} the pixel offset from the direct light position $\Delta Y$ is converted to wavelength $\lambda$ by a linear equation as
\begin{equation}
  \lambda_\mathrm{tool} [\micron] = 0.00967625 \times \Delta Y [\mathrm{pix}] + 3.12121.
  \label{eq.cal3.tool}
\end{equation}
This calibration, however, does not take into account the wavelength dependence of the refractive index of Ge, which cannot be ignored in the correction for the second-order light contamination.

The field-of-view (FOV) of the NIR channel consists of two slits (Ns and Nh) and two square-shaped apertures (Nc and Np, see figure 2 of Paper I).
The two square apertures are intended for slit-less spectroscopy of point-like objects.
This concept is briefly explained in section \ref{sec.cal3.method}.
The wavelength calibration should be determined from the slit spectroscopy, where the source position in the observed image is fixed.
As in Paper I, we measured the difference in the toolkit's wavelength calibration from observations of emission-line objects carried out in the Ns slit (5\arcsec\ width $\times$ 0\farcm8 length).
The AKARI Astronomical Observation Template (AOT) of the used observations was all of IRCZ4 b;Ns, where ``IRC'', ``Z'', ``4'', ``b'', and ``Ns'' mean the instrument, Phase 3, spectroscopic mode, grism mode, and FOV section, respectively.
The observation sequence of IRCZ4 is almost the same as that of the corresponding AOT in Phases 1 and 2 (IRC04), which is described in detail in \citet{OhyamaEtAl_2007e} and \citet{OnakaEtAl_2007}.
Although Paper I used planetary nebulae as calibrators, here instead we employed Galactic H\,\emissiontype{II} regions observed in the director's time for the IRC team (proposal ID: DTIRC) because there were fewer observations of planetary nebulae in Phase 3.
As a result, 10, 10, and 21 observations of different 10, 10, and 13 H\,\emissiontype{II} regions were obtained in Phases 3-1, 3-2, and 3-3, respectively. 
The observation dates are plotted in figure \ref{fig.cal3.temp_time}.
Out of the total 41 observations, 14 were also analyzed by \citet{MoriEtAl_2014} for scientific purposes.

\begin{figure}
  \begin{center}
    \includegraphics[width=8cm]{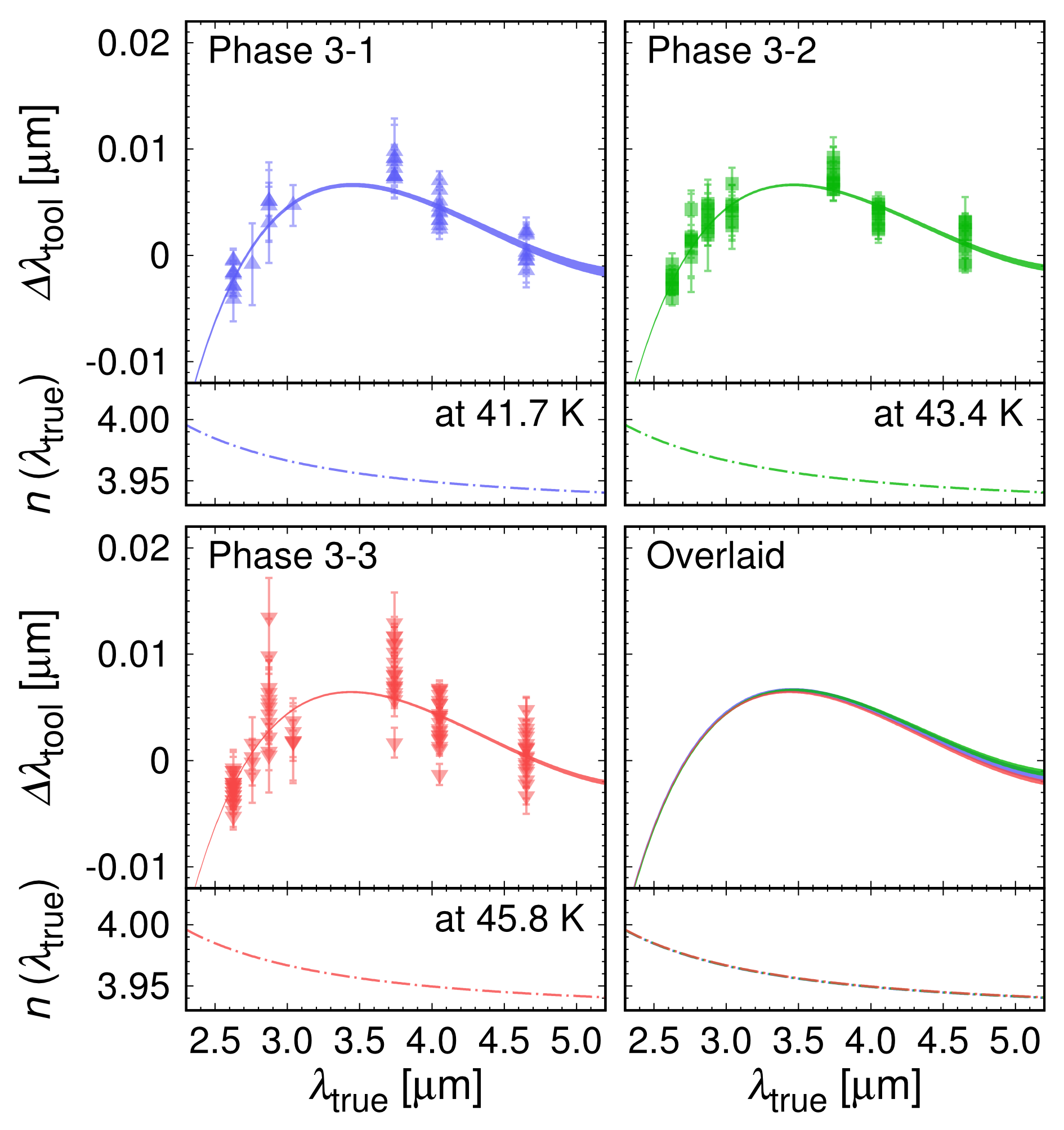}
  \end{center}
  \caption{%
    The measured differences of the previous wavelength calibration and the refractive index of Ge for the three sub-phases are shown in the quadrants other than the right bottom one.
    The upper panel in each quadrant shows the measured wavelength difference $\Delta\lambda_\mathrm{tool}\equiv\lambda_\mathrm{tool}-\lambda_\mathrm{true}$, where $\lambda_\mathrm{tool}$ is the central wavelength obtained in the line fitting, and $\lambda_\mathrm{true}$ the theoretical value.
    The filled curve represents the $1\sigma$ uncertainty in the fitted model (see text).
    The lower panel of each quadrant shows the refractive index of Ge at the average detector temperature among the observations used for the measurement of $\lambda_\mathrm{tool}$.
    In the right bottom quadrant, the model uncertainties and the refractive indices in the three sub-phases are overlaid, but in the lower panel the differences among the three lines are too small and can be hardly seen.}
  \label{fig.cal3.tool-true}
\end{figure}

From each observation, a 2D spectral image was extracted by the toolkit in the standard manner.
The image usually contained many hot pixels, which were treated as missing values in the toolkit.
We filled the missing pixels with the median of their adjacent valid pixels.
We extracted a 1D spectrum from the filled image by summing along the spatial axis with a width of 25 pixels ($\sim$36\arcsec).
This width was determined to avoid leak light from the adjacent two square-shaped apertures to the slit.
We then fitted the H\,\textsc{i} Br$\alpha$, Br$\beta$, Pf$\beta$, Pf$\gamma$, Pf$\varepsilon$, Pf11, and Pf12 lines in the spectrum with a Gaussian on a linear baseline with the line width fixed to a full width at half maximum (FWHM) of 0.031~\micron.
This is the same value adopted by \citet{MoriEtAl_2014} and was set from the physical width of the slit and the wavelength dispersion of the grism.
We did not fit the Pf$\delta$ line because it was overlapped by the emission from polycyclic aromatic hydrocarbons at 3.3~\micron.
The central wavelength $\lambda_\mathrm{tool}$ obtained from the line fitting to the spectra calibrated upon equation (\ref{eq.cal3.tool}) is plotted as the difference from the theoretical value $\lambda_\mathrm{true}$ ($\Delta\lambda_\mathrm{tool}\equiv\lambda_\mathrm{tool}-\lambda_\mathrm{true}$), in the three quadrants, other than the right bottom one, of figure \ref{fig.cal3.tool-true}.

As in Paper I, we fitted the wavelength difference $\Delta\lambda_\mathrm{tool}$ measured in each sub-phase, taking into account the refractive index of Ge $n$ as a function of $\lambda$.
Since the temperature dependence of $n$ in 41--47~K is small ($<$0.4\%) at 1.2--5.5~\micron\ \citep{FreyEtAl_2006f}, the results in the three sub-phases are expected to be the same unless other causes affect them.
The relation between $\lambda$ and $\Delta Y$ is expressed as
\begin{eqnarray}
  && m \lambda = d \left[ n(\lambda) \sin\alpha - \sin (\alpha - \theta) \right], \label{eq.cal3.interfere} \\
  && \Delta Y= (L/p)\tan\theta, \label{eq.cal3.Delta-Y}
\end{eqnarray}
where $m$ is the order number, $d$ and $\alpha$ are the groove spacing and blaze angle of the grism, respectively, $\theta$ is the emitting angle from the grism, $L$ is the interval from the final focus lens to the detector, and $p$ is the pixel pitch of the detector (see figure 1 of Paper I).
When $\theta$ is small and $n$ is treated as constant, these equations become a linear relation:
\begin{equation}
  \Delta Y = \frac{L}{p} \left[ \frac{m\lambda}{d} - (n - 1) \alpha \right].
  \label{eq.cal3.linear}
\end{equation}
The adoption of equation (\ref{eq.cal3.tool}) in the toolkit was made on this assumption.
The difference $\Delta\lambda_\mathrm{tool}$ was fitted with equations (\ref{eq.cal3.tool}), (\ref{eq.cal3.interfere}), and (\ref{eq.cal3.Delta-Y}).
In the fitting, $\alpha$ and $L$ were taken as free based on the assumption that their nominal values of 2\fdg86 and 61.9~mm had large uncertainties.
The refractive index $n(\lambda)$ was calculated from the equation of \citet{FreyEtAl_2006f} at the detector temperature averaged over the observations used for the line fitting.
The other parameters, $d$ and $p$, were fixed to their design values of 21~\micron\ and 30~\micron~pix$^{-1}$, respectively.

The resultant values of $\alpha$ and $L$ for the three sub-phases are tabulated in table \ref{tab.cal3.alphaL}, and the fitted lines of $\Delta\lambda_\mathrm{tool}$ are shown in figure \ref{fig.cal3.tool-true} along with $n(\lambda)$.
The uncertainty in the fitted model were estimated by error propagation from those in the measured $\Delta\lambda_\mathrm{tool}$.
The curves for the three sub-phases coincided with each other within the $1\sigma$ uncertainties .
Therefore, we did not find any significant temperature dependence in our wavelength calibration.
Then we fitted the equations in the same manner to all the data points without dividing them into sub-phases.
The values obtained for the two parameters are also listed in table \ref{tab.cal3.alphaL}.

\begin{table}
  \tbl{Average detector temperature and the fitted value of $\alpha$ and $L$.}{%
  \begin{tabular}{cccc}
    \hline
    Phase & $T$ (K) & $\alpha$ (deg)      & $L$ (mm)         \\
    \hline
    1, 2  & $\sim$6 & $2.8690 \pm0.0003 $ & $63.92 \pm0.03 $ \\
    \hline
    3-1   & 41.7    & $2.86966\pm0.00018$ & $63.894\pm0.013$ \\
    3-2   & 43.4    & $2.86957\pm0.00011$ & $63.908\pm0.009$ \\
    3-3   & 45.8    & $2.86938\pm0.00009$ & $63.881\pm0.007$ \\
    All   & 44.3    & $2.86952\pm0.00007$ & $63.891\pm0.005$ \\
    \hline
  \end{tabular}}\label{tab.cal3.alphaL}
\end{table}

\begin{figure}
  \begin{center}
    \includegraphics[width=8cm]{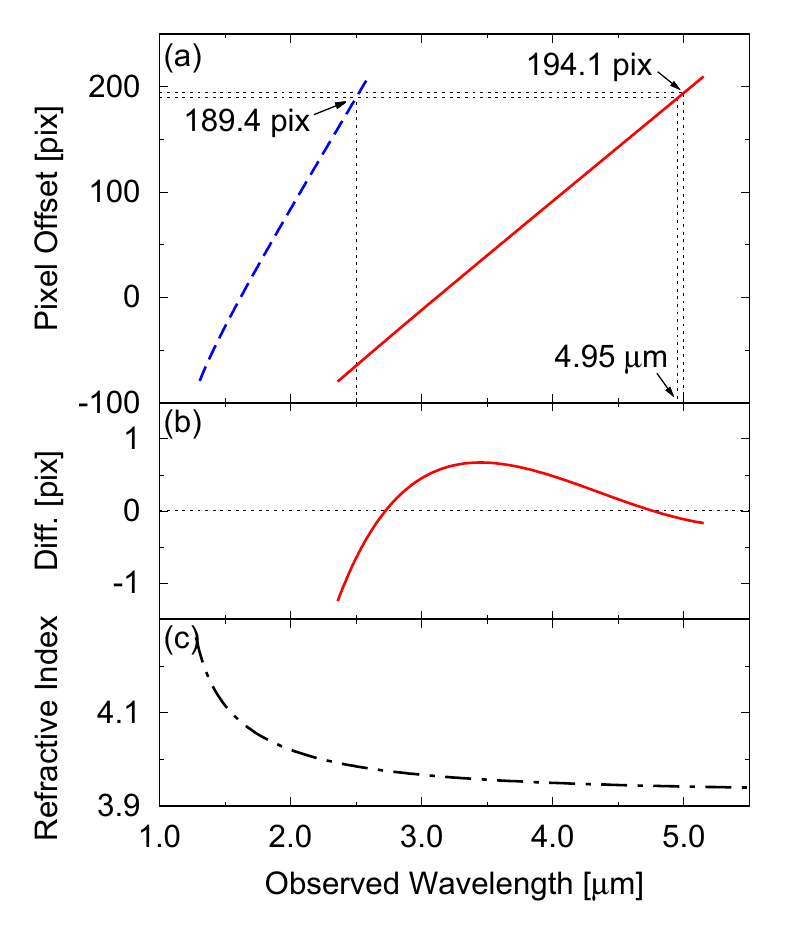}
  \end{center}
  \caption{%
    Revised wavelength calibration curve for Phase 3.
    The red solid and blue dashed lines represent relations for the first- and second-order light, respectively.
    (a) Relation between the pixel offset and the wavelength for the first- and second-order light.
    (b) Difference between our and the toolkit's wavelength calibration for the first-order light.
    The change from the value calculated by the toolkit to our value is shown.
    (c) Refractive index of Ge at 44.3~K.}
  \label{fig.cal3.wave-pos}
\end{figure}

Using the model fitted to the entire observations, we calculated the relation between the pixel offset $\Delta Y$ and the first- and second-order wavelengths $\lambda^{(1)}$ and $\lambda^{(2)}$.
This relation appears in figure \ref{fig.cal3.wave-pos}(a) and indicates that the second-order light of $\lambda^{(2)}=2.5~\micron$ is incident at the same position as the first-order light of $\lambda^{(1)}=4.95~\micron$.
This clarifies the presence of the contamination.
Even if the order-sorting filter coated onto the grism perfectly cuts off wavelengths shorter than 2.5~\micron, the second-order light overlaps with the first order spectrum in the range $\lambda^{(1)}>4.95~\micron$ due to the nonlinearity between $\lambda$ and $\Delta Y$ caused by the wavelength dependence of $n(\lambda)$.
The uncertainty of our wavelength calibration for the first-order light is 0.0014~\micron\ (0.14~pixels) and 0.0019~\micron\ (0.19~pixels) at the shorter and longer wavelength sides, respectively, being estimated from the root mean square of the residuals of the final fit to $\Delta\lambda_\mathrm{tool}$.
These uncertainties are smaller than that in the initial linear wavelength calibration (0.5~pixels; \cite{OhyamaEtAl_2007e}) because our model successfully represented the curvature of $\Delta\lambda_\mathrm{tool}$.

\section{Spectral Response Calibration}
\label{sec.cal3.res}

For each sub-phase, as in Paper I, we derived the spectral response functions for the first- and second-order light.
In this section, we explain the response calibration, the selection of the standard objects used for the calibration, the derivation of the spectral response functions, and the evaluation of the temperature dependence.

\subsection{Calibration Method}
\label{sec.cal3.method}

Objects that allow us spectral calibration by having bright featureless spectra were only observed in the two square-shaped apertures, not in the slits.
In these apertures, slit-less spectroscopy was performed, where the object position in spectroscopic frames was determined from a reference imaging frame obtained sequentially with them by rotating the filter wheel.
The two apertures, Nc and Np, are sized in 10\arcmin$\times$9.5\arcmin\ and 1\arcmin$\times$1\arcmin, respectively.
While the larger aperture Nc is aimed for multi-object spectroscopy, the smaller one Np is intended for single-target spectroscopy that avoids confusion with other sources lying along the dispersion direction.
We thus used observations carried out in Np (AOT IRCZ4 b;Np) for the following response calibration as in Paper I.

If there is no second-order light contamination, the raw output of the $i$th pixel, after the linearity of the detector had been corrected, can be written as $N_i=R(\lambda_i)F_\nu(\lambda_i)$, where $\lambda_i$ is the wavelength that enters the pixel of interest, $R$ is the spectral response function of the system, and $F_\nu$ is the flux density observed.
In this case, the response can be obtained with one standard object of known $F_\nu$.
However, if the pixel is contaminated by second-order light, $N_i$ becomes the sum of the first- and second-order components:
\begin{equation}
  N_i = R^{(1)}(\lambda_i^{(1)})F_\nu(\lambda_i^{(1)}) + R^{(2)}(\lambda_i^{(2)})F_\nu(\lambda_i^{(2)}).
\end{equation}
The first- and second-order responses, $R^{(1)}$ and $R^{(2)}$ are indecomposable unless two different objects are prepared.

As in Paper I, we adopted two types of standard objects: (1) active galactic nuclei (AGNs) whose near-IR spectra do not show any emission or absorption and thus can be modeled easily and (2) flux standard stars whose spectra are already templated.
They are red and blue, respectively, and produce contrasting amounts of contamination, which facilitates disentangling the two orders.
The two types of objects are described in the following two sections.

Once $R^{(1)}$ and $R^{(2)}$ are obtained, the second-order light contamination can be subtracted by extending the above relation between $N_i$ and $F_\nu(\lambda_i)$ to a matrix form: $\boldsymbol{N}=${\bf\sf R}$\boldsymbol{F_\nu}$.
Here, $\boldsymbol{N}$ and $\boldsymbol{F_\nu}$ are vectors of $N_i$ and $F_\nu(\lambda_i)$, and {\bf\sf R} is a matrix in which $R^{(1)}$ and $R^{(2)}$ are the diagonal and off-diagonal elements, respectively.
The inverse matrix {\bf\sf R}$^{-1}$ can be found easily.
See Paper I for more details.

\subsection{Red Standard AGNs}

From earlier published results for AKARI spectroscopy, we searched for AGNs for which we could build a reliable model spectrum.
First, we selected 41 type-I AGNs for which \citet{KimEtAl_2015} had not identified any emission and absorption features.
Second, we found 29 AGNs that did not have significant polycyclic aromatic hydrocarbon emission (3.3~\micron), Br$\alpha$ emission (4.05~\micron), or H$_2$O ice absorption (3.1~\micron) in the investigations conducted by \citet{ImanishiEtAl_2010c} and \citet{YamadaEtAl_2013}.
To create a model spectrum of a red object, we needed a Spitzer/infrared spectrograph (IRS) 5.2--32~\micron\ spectrum and reliable 2MASS and WISE magnitudes.
Among the selected AGNs, 41 had IRS observations.
Relying on the 2MASS catalog, we restricted the objects to those of the best photometric quality (\verb|ph_qual|=AAA), unaffected by known artifacts (\verb|cc_flg|=000),\footnote{For column descriptions, see Explanatory Supplement to the 2MASS All Sky Data Release and Extended Mission Products (https://www.ipac.caltech.edu/2mass/releases/allsky/doc/sec2\_2a.html).} brighter than 13 mag in the $K_s$ band, and having no neighbors brighter than 1\% of its brightness in at least one band within 15\arcsec.
We then selected objects that also have the best photometric quality in the WISE catalog (\verb|ph_qual|=AAAA).\footnote{For a description of the quality flag, see Explanatory Supplement to the
AllWISE Data Release Products (http://wise2.ipac.caltech.edu/docs/release/allwise/expsup/sec2\_1a.html).}
Finally, 23 AGNs satisfied these criteria.

\begin{figure}
  \begin{center}
    \includegraphics[width=8cm]{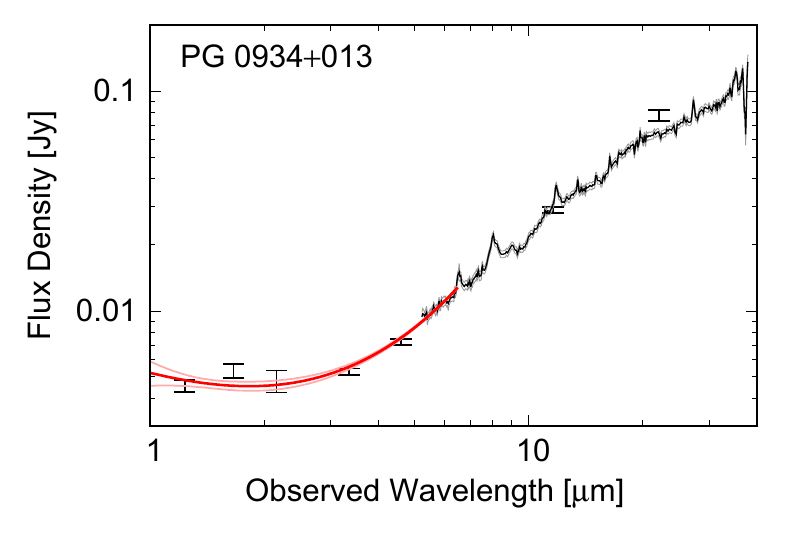}
  \end{center}
  \caption{%
    An example of the model spectra of red standard AGNs.
    The points with the error bars represent the flux densities derived from the 2MASS and WISE magnitudes with color correction.
    The black solid line and the accompanying gray lines denote the Spitzer/IRS spectrum and its uncertainty, respectively.
    The IRS spectrum is rescaled so that it fits the WISE $W3$ and $W4$ bands.
    The red solid line and the accompanying light red lines indicate the fitted cubic function and $1\sigma$ uncertainty calculated by error propagation, respectively, which were used as the model spectrum in the calibration.
    The goodness of fit is $\chi_\nu^2\equiv\chi^2/\mathrm{dof}=1.06$.}
  \label{fig.cal3.ULIRG}
\end{figure}

The 2MASS and WISE magnitudes of each AGN were converted into flux densities in janskys based on the zero magnitudes presented by \citet{CohenEtAl_2003} and \citet{JarrettEtAl_2011a}.
The color correction was performed using the spectral slope inferred from the original band fluxes.
The calibrated IRS spectrum was downloaded from the IRS Enhanced Products (version S18.18.0) in the NASA/IRAC Infrared Science Archive.
The spectrum was then rescaled so that it matched the W3 and W4 fluxes because the IRS slit spectroscopy could miss flux compared to the AKARI slitless observation.
We then fitted the $J$, $H$, $K_s$, $W1$, and $W2$ band fluxes and the IRS data points at wavelengths shorter than 6.5~\micron\ with a cubic function on the $\log F_\nu$ versus $\log\lambda$ plane.
The degrees of freedom (dof) in this configuration were 44.
We rejected bad fits that had a reduced $\chi^2$ value $\chi_\nu^2\equiv\chi^2/\mathrm{dof}$ apart from unity ($\chi_\nu^2<0.63$ or $\chi_\nu^2>1.46$, corresponding to the outside of the 95\% confidence range).
Eventually, 13 AGNs were accepted as red standard objects: PG0934$+$013, PG0947$+$396, PG1049$-$005, PG1202$+$281, PG1244$+$026, PG1415$+$451, PG1501$+$106, PG1519$+$226, PG1545$+$210, PG1700$+$518, PG1704$+$608, IRAS 23060$+$0505, and J1353157$+$634545.
An example of the model spectra is shown in figure \ref{fig.cal3.ULIRG}.
The number of observations of the 13 AGNs in Phases 3-1, 3-2, and 3-3 was 11, 4, and 19, respectively.
Their observation dates are plotted in figure \ref{fig.cal3.temp_time}.
Some AGNs were observed in more than two sub-phases.
Finally, Phases 3-1, 3-2, and 3-3 contained five, three, and eight objects, respectively.

A 2D spectral image was obtained from each observation using the toolkit.
Bad pixels were filled as described in section \ref{sec.cal3.wave}.
From the corrected 2D image, a raw 1D spectrum expressed in terms of the detector output $N$ versus the pixel offset $\Delta Y$ was extracted with a spatial width of 5 pixels (7\farcs3).
This width corresponds to the FWHM of the point spread function (PSF) in the spectroscopic mode\footnote{Section 4.4 of AKARI IRC Data User Manual for Post-Helium (Phase 3) Mission (http://www.ir.isas.jaxa.jp/AKARI/Observation/support/IRC/IDUM/ IRC\_IDUM\_P3\_1.1.pdf).} and gives an optimal signal-to-noise ratio (S/N).
The signal missed by the adoption of the finite extraction width was corrected for in the toolkit's processes estimated from the PSF.
This aperture correction is specifically adjusted for Phase 3 because the PSF size in Phase 3 is slightly degraded from that in Phases 1 and 2 \citep{OnakaEtAl_2010}.

We stacked multiple raw spectra of the same AGN in the same sub-phase to reduce the uncertainty except for IRAS 23060$+$0505 in Phase 3-2.
Although this galaxy had been observed twice during Phase 3-2, we regarded the two observations as independent and created four AGN spectra in this sub-phase, to make the number equal to that of blue standard stars observed in this sub-phase.
The motivation making the equal number of spectra is to try to see the variation of the response within each sub-phase by comparing the values calculated from spectrum pairs obtained at lower and higher temperatures.
At last, five, four, and eight AGN raw spectra were obtained for Phases 3-1, 3-2, and 3-3, respectively.
These spectra were paired with those of the standard stars (see also the next section).

\subsection{Blue Standard Stars}

Several K- and A-type standard stars were observed many times for calibration throughout Phase 3 in the director's time of the IRC team.
In particular, the K0 III star KF09T1 (2MASS J17592304$+$6602561) was monitored frequently.
We chose six bright stars (KF09T1, KF01T4, KF03T1, KF03T2, KF03T4, and TYC 4212-455-1) as the blue standard objects.
The spectra of these stars were templated by M. Cohen and his collaborators \citep{CohenEtAl_1996f,CohenEtAl_1999d,CohenEtAl_2003a,CohenEtAl_2003,Cohen_2003}.
We used these templates as the model spectra for our calibration.
The number of observations of the chosen standard stars during Phases 3-1, 3-2, and 3-3 was 41, 45, and 123, respectively.
Their observation dates are plotted in figure \ref{fig.cal3.temp_time}.
In the same way as we did for the AGNs, 1D raw spectra of the stars were extracted and then stacked.
The stacking for KF09T1 in each sub-phase was performed separately for several subsets so that the number of stacked spectra was the same for the AGNs.
The subsets were defined by equally dividing the observations ordered by their detector temperatures to investigate possible temperature variation in the sub-phase.
We determined the pairs of the stacked spectra of the stars and the AGNs according to the average temperature of observations of the stacked spectra.

\subsection{Temperature Dependence of Response Functions}

\begin{figure*}
  \begin{center}
    \begin{tabular}{@{}c@{}c@{}}
      \includegraphics{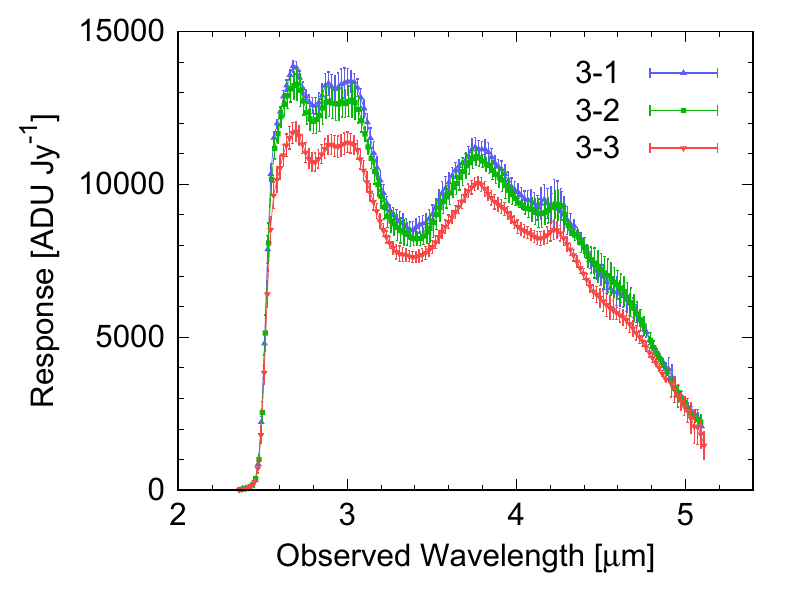} &
      \includegraphics{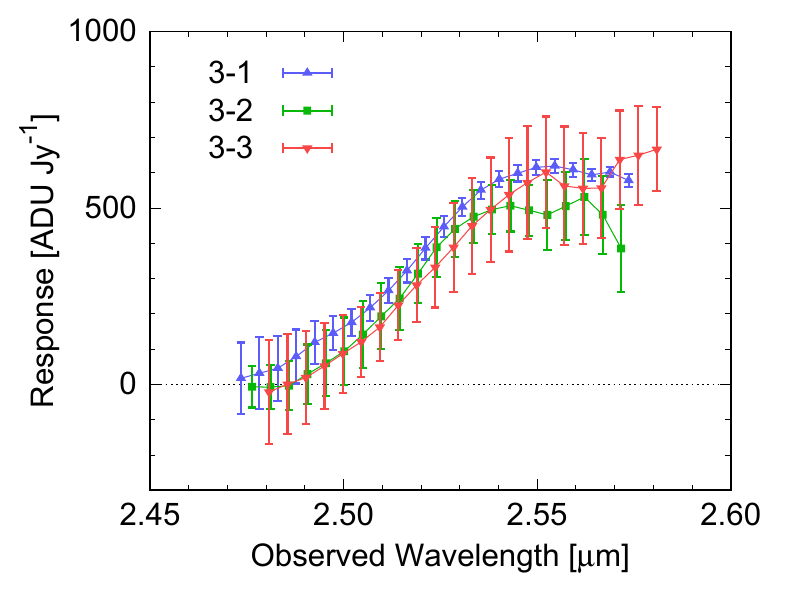} 
    \end{tabular}
  \end{center}
  \caption{%
    Spectral response curves for the three sub-phases.
    Left and right panels show the response for the first- and second-order light, respectively.}
  \label{fig.cal3.res123}
\end{figure*}

From the wavelength calibration, we can be certain that the second-order light contamination exists only for first-order wavelengths ($\lambda^{(1)}$) longer than 4.9~\micron.
For $\lambda^{(1)}>4.9~\micron$, we calculated the spectral response functions for the first- and second-order light, $R^{(1)}$ and $R^{(2)}$, from each standard object pair in the same way as we did in Paper I.
For $\lambda^{(1)}<4.9~\micron$, on the assumption that $R^{(2)}=0$, we derived $R^{(1)}$ only from the stars assuming that the Cohen templates are better established than our AGN model spectra through other calibrations, such as 2MASS and Spitzer/IRS.
The response variation within a sub-phase could not be clearly seen because the variation among the subsets was small and the S/N of each response curve was not sufficient.
To investigate the variation among the three sub-phases, we averaged the response curves obtained in each sub-phase after 5-pixel smoothing and estimated the uncertainty from the standard deviation of the curves.
The results are presented in figure \ref{fig.cal3.res123}.
In Phase 3-1, $R^{(2)}$ showed better S/N than in the other two sub-phases because the detector was most sensitive due to the low temperature and most stable due to the small temperature variation among the observations used for the calibration.

\begin{figure}
  \begin{center}
    \includegraphics{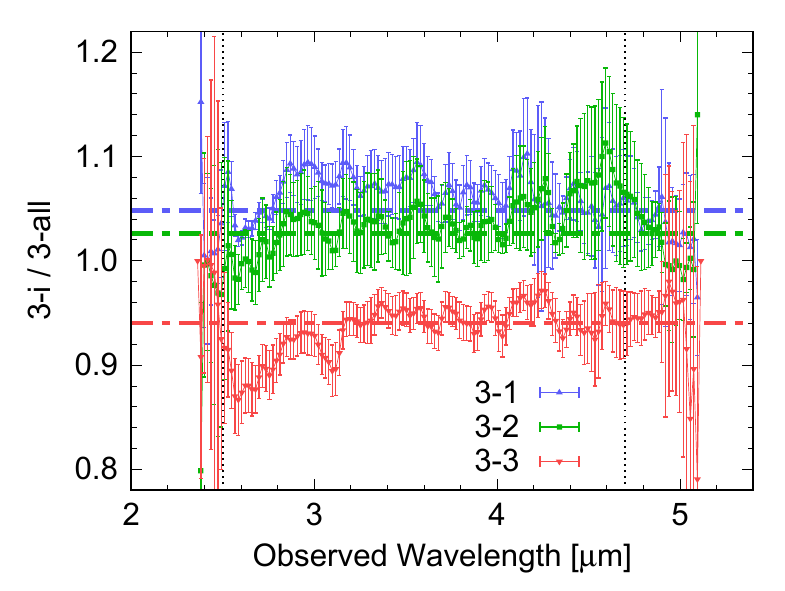}
  \end{center}
  \caption{%
    Ratio of the response in each sub-phase to the average of all sub-phases in Phase 3.
    The thick dot-dashed lines indicate the level of the response ratio averaged in the 2.5--4.7~\micron\ range.}
  \label{fig.cal3.ratio}
\end{figure}

We cannot see the temperature dependence of $R^{(2)}$ seen in the right panel of figure \ref{fig.cal3.res123} because the results from the three sub-phases have a large uncertainty.
In contrast, the left panel clearly shows that $R^{(1)}$ decreased during Phase 3.
The decrease does not appear to have an obvious wavelength dependence.
To confirm this, we averaged the three response curves and derived the ratio of each response to the average.
The ratios obtained are shown in figure \ref{fig.cal3.ratio}.
The response ratios do not show a clear wavelength dependence although they have large uncertainties around the edges of the wavelength coverage. 
In each sub-phase, the level of the response ratio averaged in 2.5--4.7~\micron\ hits most of the data points and their error bars, as shown in figure \ref{fig.cal3.ratio}.
We, thus, assume that the decrease of $R^{(1)}(\lambda)$ has no wavelength dependence.

\begin{figure}
  \begin{center}
    \includegraphics{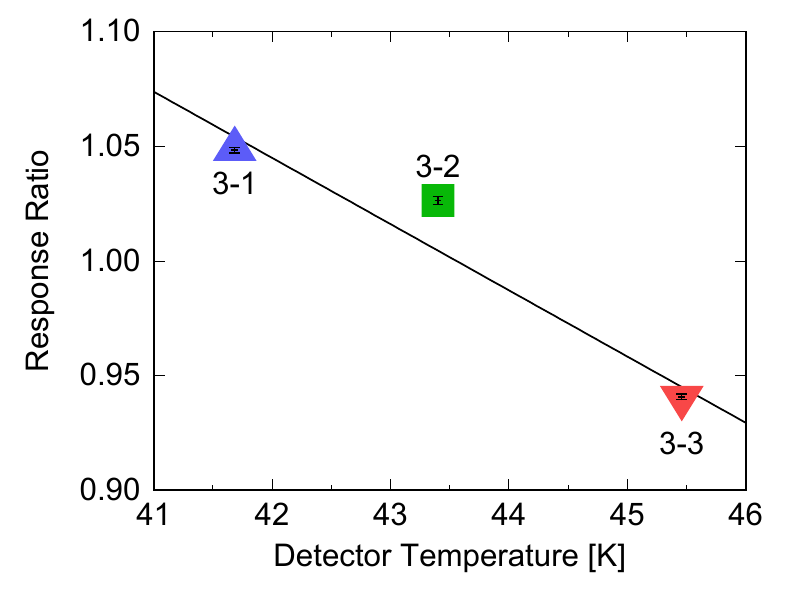}
  \end{center}
  \caption{%
    Plot of the averaged response ratio against the average detector temperature of the observations used for the calibration.  
    The black solid line is the linear function fitted to the three points and is the temperature correction factor in equations (\ref{eq.cal3.Tcor}), (\ref{eq.cal3.param-a}), and (\ref{eq.cal3.param-T}). The error bar at each point represents the uncertainty in the mean response ratio propagated from those shown in figure \ref{fig.cal3.ratio}.}
  \label{fig.cal3.Tvar}
\end{figure}

Figure \ref{fig.cal3.Tvar} plots the level of the averaged response ratio against the average detector temperature of the observations used for the spectral response calibration.
To derive a correction factor for the response decrease, we approximated it as a linear relation and fitted a function of the form
\begin{equation}
  f(T)=1+a(T-T_0),
  \label{eq.cal3.Tcor}
\end{equation}
to the three points in figure \ref{fig.cal3.Tvar}.
As a result,
\begin{equation}
  a   = -0.0290~\mathrm{K^{-1}},
  \label{eq.cal3.param-a}
\end{equation}
and
\begin{equation}
  T_0 = 43.55~\mathrm{K}.
  \label{eq.cal3.param-T}
\end{equation}
were obtained.
The decreasing slope of the response found here from spectroscopic observations coincides with the value derived from NIR photometric observations (\cite{OnakaEtAl_2010}; Yamashita et al. in preparation).
Thus, the decline in the response can be attributed to the parts of the NIR channel shared in the spectroscopic and photometric modes.
Because it is unlikely that the transmittance of the Si lenses changed, the dominant cause of the decline is probably due to a decrease of the detector sensitivity.

Users who wish to reduce the data from a spectroscopic observation can adopt the averaged spectral response curve (in units of ADU/Jy) scaled by the factor $f(T)$.
Although the temperature dependence of the spectral response for the second-order light is not clear, we assume that overall it scales with the same factor $f(T)$.
This assumption should not cause any practical problems because the change in $f(T)$ during Phase 3 ($\sim10\%$) is as small as the uncertainty of $R^{(2)}(\lambda)$ in each sub-phase. 

\section{Comparison with the Calibration in Phases 1 and 2}
\label{sec.cal3.comp}

In the wavelength calibration, the values of $\alpha$ and $L$ obtained in Phases 1 and 2 and Phases 3-1, 3-2, and 3-3 coincide well with each other (see table \ref{tab.cal3.alphaL}), and the difference in the wavelength calibration curves for the former and latter phases is about 0.2~pixels.
Thus there is little practical difference in the wavelength calibration for Phase 3, compared to that for Phases 1 and 2.

\begin{figure*}
  \begin{center}
    \begin{tabular}{@{}c@{}c@{}}
      \includegraphics[width=8cm]{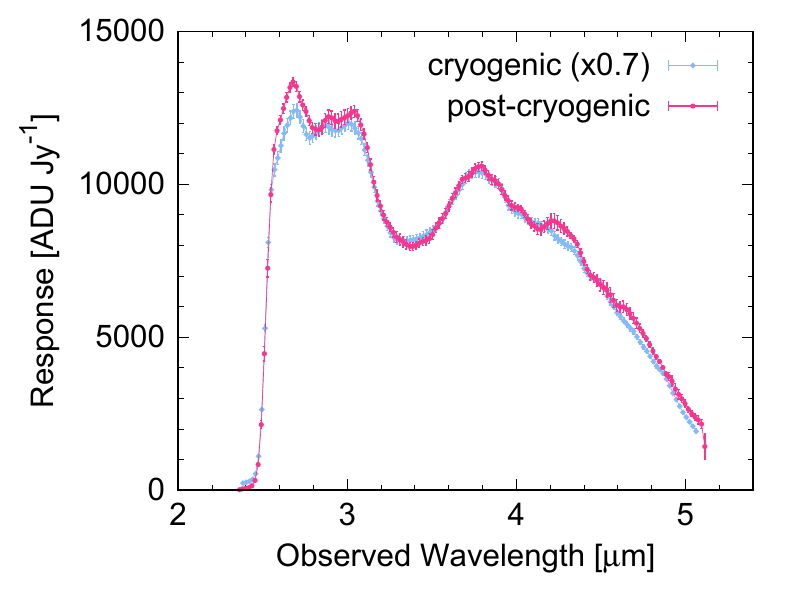} &
      \includegraphics[width=8cm]{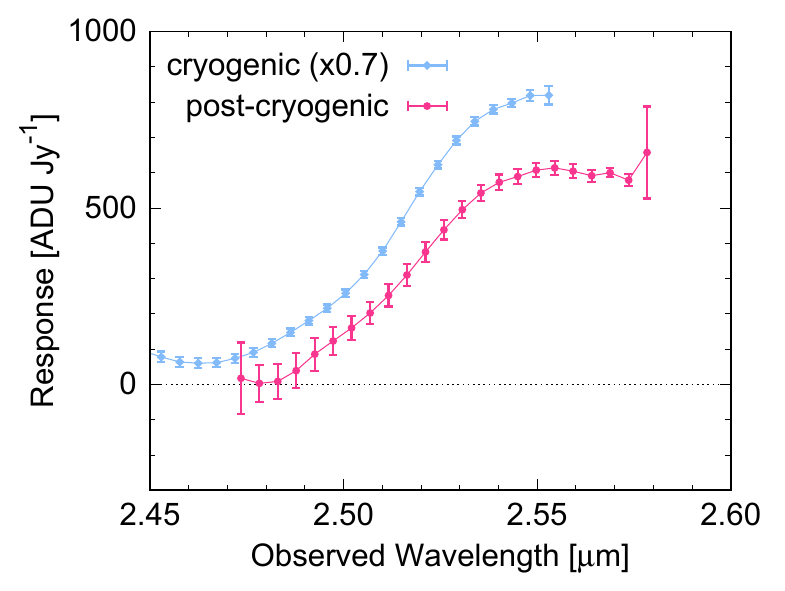} 
    \end{tabular}
  \end{center}
  \caption{%
    Comparison of the spectral responses in Phase 3 with those in Phases 1 and 2.
    The response curves for Phases 1 and 2 obtained in Paper I are multiplied by a factor of 0.7.
    Left: $R^{(1)}(\lambda)$.
    Right: $R^{(2)}(\lambda)$.}
  \label{fig.cal3.resall}
\end{figure*}

Figure \ref{fig.cal3.resall} shows $R^{(1)}(\lambda)$ and $R^{(2)}(\lambda)$ averaged among the three sub-phases and compares them with those for Phases 1 and 2 obtained in Paper I.
The relative shape of $R^{(1)}(\lambda)$ does not change from Phases 1 and 2 to Phase 3.
This comparison also supports the assumption in the derivation of $f(T)$ that the decrease of the response with temperature does not have a wavelength dependence.
The degradation in $R^{(1)}$ from Phases 1 and 2 to 3 is estimated to be 0.7.
This value is consistent with that reported earlier \citep{OnakaEtAl_2010}.

The right panel of figure \ref{fig.cal3.resall} is the same comparison as the left one but for $R^{(2)}(\lambda)$.
Compared to $R^{(1)}$, $R^{(2)}$ decreased by more than a factor of 0.7 from Phases 1 and 2.
It decreased by a factor of 0.52.
Thus, the strength of the second-order light contamination relative to that of the first-order light in Phase 3 is lower than that in Phases 1 and 2 by 25\% when we observe the same object.
The origin of this decline might be due to the optics because it seems unnatural that the detector sensitivity drops specifically at a certain wavelength.
Although the detail is unclear at present, as the contamination is successfully corrected for in practice, we do not further pursue the origin in this paper.

\section{Summary}
\label{sec.cal3.sum}

We investigated the second-order light contamination in the AKARI NIR grism spectroscopy specifically for the post-cryogenic phase (Phase 3).
We performed wavelength and spectral response calibrations like those conducted in Paper I for the cryogenic phase (Phases 1 and 2).
To assess the temperature dependence of the wavelength and spectral response calibrations, we defined three sub-phases corresponding to different ranges of the detector temperature.
The wavelength calibrations for the three sub-phases coincided with each other and did not show any temperature dependence.
The final wavelength calibration curve obtained from the measurement of the emission lines throughout Phase 3 confirmed the presence of the second-order light contamination, as in Phases 1 and 2.
We showed that the spectral response of the first-order light decreased by $\sim$10\% from the beginning to the end of Phase 3.
Based on the approximation that the decline of the response linearly relates to the temperature, a correction factor for the temperature dependence was obtained.
The relative strength of the second-order light contamination to the first-order light was found to be 25\% smaller than that in Phases 1 and 2.
We introduced our correction procedure into the official data reduction toolkit.
The updated version of the toolkit is now in preparation for public distribution.

\begin{ack}
We acknowledge the anonymous referee for his or her careful reading and helpful comments that improved the quality of this paper.
AKARI is a JAXA project with the participation of ESA.
We thank all the members of the AKARI project for their continuous help and support.
This publication makes use of data products from the Two Micron All Sky Survey, which is a joint project of the University of Massachusetts and the Infrared Processing and Analysis Center/California Institute of Technology, funded by the National Aeronautics and Space Administration and the National Science Foundation.
This publication makes use of data products from the Wide-field Infrared Survey Explorer, which is a joint project of the University of California, Los Angeles, and the Jet Propulsion Laboratory/California Institute of Technology, funded by the National Aeronautics and Space Administration.
This work is based in part on observations made with the Spitzer Space Telescope, which is operated by the Jet Propulsion Laboratory, California Institute of Technology under a contract with NASA.
This research has made use of the NASA/ IPAC Infrared Science Archive, which is operated by the Jet Propulsion Laboratory, California Institute of Technology, under contract with the National Aeronautics and Space Administration.
This study is supported by JSPS KAKENHI Grant Number JP17J01789.
\end{ack}


\end{document}